\documentclass[11pt,a4paper]{article}
\usepackage[a4paper]{geometry}
\RequirePackage[numbers]{natbib}
\usepackage{amsmath,amsfonts,amssymb,amsthm}
\usepackage{graphicx}
\usepackage{enumerate}
\usepackage{natbib}
\usepackage{fancyhdr}
\usepackage{cancel}
\usepackage[hyperindex,breaklinks]{hyperref}
\hypersetup{frenchlinks=true}
\hypersetup{
  colorlinks   = true, 
  urlcolor     = blue, 
  linkcolor    = blue, 
  citecolor    = black 
}

\usepackage{xspace}
\newcommand{\xmath}[1]{\ensuremath{#1}\xspace}
\newcommand{\E}{\xmath{\mathbb{E}}}
\newcommand{\Prob}{\xmath{\mathbb{P}}}

\newcommand{\ex}{\xmath{\mathrm{e}}}

\newcommand{\cA}{\xmath{\mathcal{A}}}
\newcommand{\cL}{\xmath{\mathcal{L}}}

\newcommand{\fb}{\xmath{\bar{f}}}
\newcommand{\io}{\xmath{\mathrm{i}}}
\newcommand{\del}{\xmath{\partial}}
\newcommand{\e}{\xmath{\epsilon}}

\begin{document}

\title{Asymptotic expansion for characteristic function in Heston stochastic volatility model with fast mean-reverting correction}

\author{Ankush Agarwal\footnote{School of Technology and Computer Science, Tata Institute of Fundamental Research, Mumbai MA 400005; \emph{ankush@tcs.tifr.res.in}}}

\maketitle
\date{}

\begin{abstract}
In this note, we derive the characteristic function expansion for logarithm of the underlying asset price $X_{\tau}$ in corrected Heston model as proposed by Fouque and Lorig \cite{fouque2011fast}. 
\end{abstract}

\section{Introduction}
Heston model \cite{heston1993closed} has been the most widely used stochastic volatility model as it provides an explicit formula for calculating European option prices. But as noted by Fouque and Lorig \cite{fouque2011fast} , the model fails to capture the implied volatility skew in far in-the-money and out-of-the-money European options. A suggested explanation for this shortcoming is the single factor of volatility in the model which is not sufficient for describing the dynamics of the observed volatility process. 

In \cite{fouque2011fast}, the authors proposed a fast mean-reverting correction to the Heston model which still preserves the explicitness of the formula for European option price. It is demonstrated that the corrected Heston model captures the market implied volatility skew better than the original Heston model. 

In this note, we use similar singular perturbative expansion to derive a correction to the characteristic function of the logarithm of underlying asset price in the Heston model. 

\section{Multi-Scale Model and Characteristic function PDE}
Under the pricing risk-neutral probability measure $\Prob$, consider the price $S_t$ of an asset whose dynamics is described by the following system of stochastic differential equations:
\begin{eqnarray*}
%\label{eq:stochvol_model1}
dS_t &=& r S_t dt + \Sigma_t S_{t}dW^x_t,\\
\Sigma_t &=& \sqrt{Z_t}f(Y_t),\\
%\label{eq:stochvol_model2}
dY_t &=& \frac{Z_t}{\epsilon}(m - Y_t)dt + \nu\sqrt{2} \sqrt{\frac{Z_t}{\epsilon}} dW^{y}_t,\\
%\label{eq:stochvol_model3}
dY_t &=& \kappa(\theta - Z_t)dt + \sigma \sqrt{Z_t} dW^{z}_{t}.
\end{eqnarray*}
Here, $W^{x}_{t}$, $W^{y}_{t}$ and $W^z_{t}$ are one-dimensional Brownian motions with the correlation structure
\begin{eqnarray*}
d \langle W^x, W^y\rangle_t &=& \rho_{xy} dt,\\
d \langle W^x, W^z\rangle_t &=& \rho_{xz} dt,\\
d \langle W^y, W^z\rangle_t &=& \rho_{yz} dt,
\end{eqnarray*}
where the correlation coefficients $\rho_{xy}, \rho_{xz}$ and $\rho_{xy}$ are constants satisfying $\rho^2_{xy} < 1$, $\rho^2_{xz} < 1$, $\rho^2_{yz} < 1$, and $\rho^2_{xy} + \rho^2_{xz} + \rho^2_{yz} - 2\rho_{xy} \rho_{xz} \rho_{yz} < 1$ in order to ensure positive definiteness of the covariance matrix of the three Brownian motions. 
Next, we set $X_t = \log S_t$ and get the corresponding stochastic differential equation which describes the dynamics of $X_t$:
\begin{equation*}
%\label{eq:stochvol_model4}
dX_t = \Bigl(r - \frac{1}{2}f^2(Y_t)Z_t \Bigr) dt + \Sigma_t dW^x_t.
\end{equation*}
Let us define the characteristic function of logarithm of the underlying asset price at expiry as
\begin{equation*}
%\label{eq:char_func}
\psi^{\e}(t,x,y,z) := \E \bigl[ \mathrm{e}^{\io s X_T} \vert X_t = x, Y_t = y, Z_t = z \bigr]
\end{equation*}
where we have used the Markov property of $(X_t,Y_t,Z_t)$ and defined the characteristic function $\psi^{\e}(t,x,y,z)$, the superscript $\e$ denoting the dependence on the small parameter $\e$. 
%By defining $\psi(\cdot)$ as in \eqref{eq:char_func}, we have 
%$$ \psi^{\e}(t,s,x,y,z) = 1 \hspace{0.5pc} \text{and} \hspace{0.5pc} \psi^{\e}(\tau,-\io,x,y,z) = x \ex^{r \tau}. $$
Using the Feynman-Kac formula for $\psi^{\e}(t,x,y,z)$, a conditional expectation, we get the following PDE and boundary conditions: 
\begin{eqnarray}
\label{eq:main_pde}
\cL^{\e} \psi^{\e}(t,x,y,z) &=& 0,\\
\cL^{\e} &=& \frac{\del}{\del t} + \cL_{(X,Y,Z)},\\
\label{eq:bound_cond}
\psi^{\e}(T,x,y,z) &=& \ex^{\io s x}
\end{eqnarray}
where the operator $\cL_{(X,Y,Z)}$ is the infinitesimal generator of the process $(X_t, Y_t, Z_t)$:
\begin{align*}
\cL_{(X,Y,Z)} &= \Bigl(r - \frac{1}{2}f^2(y)z \Bigr)\frac{\del}{\del x} + \frac{1}{2}f^2(y)z \frac{\del^2}{\del x^2} + \rho_{xz} \sigma f(y) z \frac{\del^2}{\del x \del z} + k(\theta - z) \frac{\del }{\del z} + \frac{1}{2}\sigma^2 z \frac{\del^2}{\del z^2}\\
&+ \frac{z}{\e}\Bigl((m-y) \frac{\del}{\del y} + \nu^2 \frac{\del^2}{\del y^2} \Bigr) + \frac{z}{\sqrt{\e}}\Bigl( \rho_{yz}\sigma \nu \sqrt{2} \frac{\del^2}{\del y \del z} + \rho_{xy} \nu \sqrt{2} f(y) \frac{\del^2}{\del x \del y} \Bigr).
\end{align*}
We separate $\cL^{\e}$ into groups of like-powers of $1/\sqrt{\e}$. To this end, we define the operators $\cL_0$, $\cL_1$ and $\cL_2$ as follows:
\begin{align*}
\cL_0 &:= \nu^2\frac{\del^2}{\del y^2} + (m-y) \frac{\del}{\del y},\\
\cL_1 &:= \rho_{yz}\sigma \nu \sqrt{2} \frac{\del^2}{\del y \del z} + \rho_{xy} \nu \sqrt{2} f(y) \frac{\del^2}{\del x \del y},\\
\cL_2 &:= \frac{\del}{\del t} + \frac{1}{2}f^2(y)z \frac{\del^2}{\del x^2} + \Bigl(r - \frac{1}{2}f^2(y)z \Bigr)\frac{\del}{\del x}\\
&+ \frac{1}{2}\sigma^2 z \frac{\del^2}{\del z^2} + k(\theta - z) \frac{\del }{\del z} + \rho_{xz} \sigma f(y) z \frac{\del^2}{\del x \del z}.
\end{align*}
With these definitions, $\cL^{\e}$ is expressed as: 
\begin{equation}
\label{eq:operator_expansion}
\cL^{\e} = \frac{z}{\e}\cL_0 + \frac{z}{\sqrt{\e}}\cL_1 + \cL_2 .
\end{equation}
\section{Asymptotic analysis}
In order to achieve an approximation for the characteristic function $\psi^{\e}$, we perform a singular perturbation with respect to the small parameter $\e$, expanding the solution in powers of $\sqrt{\e}$
\begin{equation}
\label{eq:char_expansion}
\psi^{\e}(t,x,y,z) = \psi_0(t,x,y,z) + \sqrt{\e} \psi_1(t,x,y,z) + \e \psi_2(t,x,y,z) + \ldots .
\end{equation}
We plug \eqref{eq:char_expansion} and \eqref{eq:operator_expansion} into \eqref{eq:main_pde} and \eqref{eq:bound_cond}, and collect terms of equal powers of $\sqrt{\e}$.

\vspace{1.0pc}
\textbf{Order $1/ \e$ terms.} Collecting terms of order $1/\e$ we have the following PDE:
\begin{equation*}
%\label{eq:first_order}
z\cL_{0}\psi_0 = 0.
\end{equation*}
As $\cL_{0}$ takes derivatives with respect to only $y$, the first equation implies that $\psi_0$ is independent of $y$. In fact, $\cL_0$ is an infinitesimal generator and consequently zero is an eigenvalue with constant eigenfunctions. Thus, we seek $\psi_0$ of the form
\begin{equation*}
%\label{eq:first_term}
\psi_0 = \psi_0(t,x,z).
\end{equation*}

\vspace{1.0pc}

\textbf{Order $1/\sqrt{\e}$ terms.} Collecting terms of order $1/\sqrt{\epsilon}$ leads to the following PDE:
\begin{align*}
0 &= z\cL_{0}\psi_1 + z \cL_1 \psi_0,\\
%\label{eq:second_order}
&= z\cL_{0}\psi_1.
\end{align*}
where we have used the result that $\psi_{0}$ is independent of $y$. For the same reasons as $\psi_0$, $\psi_1$ is also independent of $y$. Thus, we seek $\psi_1$ of the form
\begin{equation*}
%\label{eq:second_term}
\psi_1 = \psi_1(t,x,z).
\end{equation*}

\vspace{1.0pc}

\textbf{Order $1$ terms.} Matching terms of order 1 leads to the following PDE and boundary condition:
\begin{align}
&0 = z\cL_{0}\psi_{2} + z\cL_1 \psi_1 + \cL_2 \psi_0, \nonumber\\
\label{eq:third_term}
&= z\cL_{0}\psi_{2} + \cL_2 \psi_0,\\
\label{eq:thirdorder_boundcond}
e^{\io sx} &= \psi_0(T,x,z),
\end{align}
where we have used the independence of $\psi_1$ on $y$. 

We note that $\eqref{eq:third_term}$ is a Poisson equation for $\psi_2$ with respect to the infinitesimal generator $\cL_{0}$. A solution for $\psi_2$ exists if and only if $\cL_{2}\psi_0$ is centered with respect to the invariant distribution of the diffusion whose infinitesimal generator is $\cL_{0}$. Thus, the centering condition becomes $$\langle \cL_{2}\psi_0 \rangle = 0,$$ where the angled brackets indicate taking the average of the argument with respect to the invariant distribution of the diffusion whose infinitesimal generator is $\cL_{0}$. Since $\psi_0$ does not depend on $y$, the centering condition becomes 
\begin{equation}
\label{eq:psi_pde}
\langle \cL_2 \rangle \psi_0 = 0.
\end{equation}

We note that the PDE \eqref{eq:psi_pde} along with the boundary condition \eqref{eq:thirdorder_boundcond} jointly define a Cauchy problem that $\psi_0(t,x,z)$ must satisfy. As we show later, this is equivalent to solving the PDE for characteristic function in Heston model. 

Using \eqref{eq:third_term} and the centering condition we can show
\begin{equation}
\label{eq:firstform_psi2}
\psi_2 = -\frac{1}{2} \cL_0^{-1} \Bigl( \cL_2 - \langle \cL_2 \rangle \Bigr) \psi_0.
\end{equation}

\vspace{1.0pc}

\textbf{Order $\sqrt{\e}$ terms.} Collecting terms of order $\sqrt{\e}$, we obtain the following PDE and boundary condition:
\begin{align}
\label{eq:fourth_term}
0 &= z\cL_{0}\psi_3 + z\cL_{1}\psi_2 + \cL_{2}\psi_{1},\\
\label{eq:fourthterm_boundcond}
0 &= \psi_1(T,x,z).
\end{align}

We note that $\psi_3$ solves the Poisson equation in $y$ with respect to $\cL_{0}$. Thus, we impose the corresponding centering condition on the source $z\cL_{1}\psi_2 + \cL_{2}\psi_{1}$, leading to 
\[ \langle \cL_{2} \rangle \psi_1 = -z \langle \cL_{1} \psi_2 \rangle.\]

Plugging $\psi$, given by \eqref{eq:firstform_psi2} into the above equation, we get 
\begin{align}
\label{eq:pde_psi1}
\langle \cL_{2} \rangle \psi_1 &= \cA \psi_0,\\
%\label{eq:operatorA_firstform}
\cA &:= \left\langle z \cL_1 \frac{1}{z} \cL_o^{-1}(\cL_2 - \langle \cL_2\rangle) \right\rangle. \nonumber
\end{align}
Note that the PDE \eqref{eq:pde_psi1} and the boundary condition \eqref{eq:fourthterm_boundcond} define a Cauchy problem that $\psi_1(t,x,z)$ must satisfy.

\vspace{1.0pc}

\section{Calculation of leading order terms}
\textbf{Calculation of $\psi_0$.} Our previous analysis shows that $\psi_0$ is the solution of the following Cauchy problem
\begin{align}
\label{eq:psi0_pde}
\langle \cL_2 \rangle \psi_0 &= 0,\\
\label{eq:psi0_boundcond}
\psi_0(T,x,z) &= e^{\io sx}.
\end{align}
Like in \cite{fouque2011fast}, without loss of generality, we normalize $f$ so that $\langle f^2 \rangle = 1.$ Thus, we can write,
\begin{align}
\label{eq:average_hestonoperator}
\langle \cL_2 \rangle &= \frac{\del}{\del t} + \frac{1}{2} z \frac{\del^2}{\del x^2} + \Bigl(r - \frac{1}{2} z \Bigr)\frac{\del}{\del x} + \frac{1}{2}\sigma^2 z \frac{\del^2}{\del z^2}\nonumber \\
&+ \kappa(\theta - z) \frac{\del}{\del z} + \rho\sigma z \frac{\del^2}{\del x \del z},
\end{align}
where $\rho := \rho_{xz}\langle f \rangle.$ It is clear that $\langle \cL_2 \rangle$ in \eqref{eq:average_hestonoperator} is a Heston model operator modulated with effective correlation term $\rho_{xz}\fb$ with $\fb := \langle f \rangle$. Thus, we choose the solution $\psi_0$ of the following form 
\begin{equation*}
%\label{eq:psi0_solform}
\psi_0 = \exp(C(T-t) + zD(T-t) +\io sx).
\end{equation*}
Plugging back the solution form into the PDE \eqref{eq:psi0_pde}, we get the following equations:
\begin{align*}
%\label{first_ricatti}
0 &= \frac{\del C}{\del t} + r\io s + \kappa \theta D,\\
%\label{second_ricatti}
0 &= \frac{\del D}{\del t} - \frac{1}{2} (s^2 + \io s) + \frac{1}{2}\sigma^2 D^2 - (\kappa - \io \rho \sigma s) D
\end{align*}
subject to $C(0) =0 $ and $D(0) = 0$. These equations can be solved to provide the following expressions 
\begin{align*}
C(\tau) &= \io r s \tau + \frac{\kappa \theta}{\sigma^2}\Bigl( (\kappa - \rho \io \sigma s +d) \tau - 2 \log \Bigl( \frac{1 - g(s) e^{\tau d(s)}}{1 - g(s)}\Bigr) \Bigr),\\
D(\tau) &= \frac{\kappa - \rho \io \sigma s +d}{\sigma^2} \Bigl( \frac{1 - e^{\tau d(s)}}{1 - g(s)e^{\tau d(s)}} \Bigr),\\
d(s) &= \sqrt{(\rho \sigma \io s - \kappa)^2  + \sigma^2(\io s + s^2)},\\
g(s) &= \frac{\kappa - \rho \sigma \io s + d}{\kappa - \rho \sigma \io s - d},\\
\tau &= T-t.
\end{align*}

\vspace{1.0pc}
\textbf{Calculation of $\psi_1$.} We first evaluate the operator $\cA$.
\begin{align*}
\cA &= \left\langle z \cL_1 \frac{1}{z} \cL_0^{-1}(\cL_2 - \langle \cL_2 \rangle) \right\rangle\\
&= \left\langle z \cL_1 \frac{1}{z} \cL_0^{-1} \frac{z}{2}(f^2 - \langle f^2 \rangle) \frac{\del^2}{\del x^2} \right\rangle - \left\langle z \cL_1 \frac{1}{z} \cL_0^{-1} \frac{z}{2}(f^2 - \langle f^2 \rangle) \frac{\del}{\del x} \right\rangle\\
&+ \left\langle z \cL_1 \frac{1}{z} \cL_0^{-1} \rho_{xz} \sigma z(f - \langle f \rangle) \frac{\del^2}{\del x \del z} \right\rangle\\
&= z \left\langle \cL_1 \phi(y) \frac{\del^2}{\del x^2} \right\rangle - z \left\langle \cL_1 \phi(y) \frac{\del}{\del x} \right\rangle +  \rho_{xz} \sigma z \left\langle \cL_1 \xi(y) \frac{\del^2}{\del x \del z} \right\rangle.
\end{align*}
The functions $\phi(y)$ and $\xi(y)$ solve the following Poisson equations in $y$ with respect to the operator $\cL_0$:
\begin{align*}
%\label{eq:def_phi}
\cL_0 \phi &= \frac{1}{2} (f^2 - \langle f^2 \rangle),\\
%\label{eq:def_xi}
\cL_0 \xi  &= f - \langle f \rangle.
\end{align*}
We use the definition of $\cL_1$ to calculate the final expression for $\cA$:
\begin{align*}
\cA &= V_1 z \frac{\del^3}{\del z \del x^2} - V_1 z \frac{\del^2}{\del z \del x} + V_2 z \frac{\del^3}{\del z^2 \del x} \\
%\label{eq:final_cA}
&+ V_3 z \frac{\del^3}{\del x^3} - V_3 z \frac{\del^2}{\del x^2} + V_4 z \frac{\del^3}{\del z \del x^2},\\
%\label{eq:def_V1}
V_1 &= \rho_{yz} \sigma \nu \sqrt{2} \langle \phi^{\prime} \rangle,\\
%\label{eq:def_V2}
V_2 &= \rho_{xz} \rho_{yz}\sigma^2 \nu \sqrt{2} \langle \xi^{\prime} \rangle,\\
%\label{eq:def_V3}
V_3 &= \rho_{xy} \nu \sqrt{2} \langle f \phi^{\prime} \rangle,\\
%\label{eq:def_V4}
V_4 &= \rho_{xy} \rho_{xz} \sigma \nu \sqrt{2} \langle f \xi^{\prime} \rangle.
\end{align*}

We go back to Equation \eqref{eq:pde_psi1} and \eqref{eq:fourth_term} to get the PDE with boundary condition:
\begin{align*}
%\label{eq:correc_pde}
\langle \cL_{2} \rangle \psi_1 &= \cA \psi_0,\\
\psi_1(T,x,z) &= 0.
\end{align*}

Ansatz:
$$\psi_1 = (\kappa \theta f_0(t,s) + z f_1(t,s)) \psi_0.$$ 

We substitute our choice of ansatz into the PDE and boundary condition. After comparing the same order of $z$ terms, we can show that $f_0(t,s)$ and $f_1(t,s)$ satisfy the following system of ODEs:

\begin{align*}
\frac{\del f_0}{\del t} &= -f_1,\\
f_1(T,s) &= 0,\\
\frac{\del f_1}{\del t} &= a(t,s)f_1(t,s) + b(t,s),\\
f_0(T,s) &= 0,\\
a(t,s) &= -\sigma^2 D(t,s) + (\kappa - \rho \sigma \io s),\\
b(t,s) &= -\bigl( V_1D (s^2 + \io s ) - V_2 \io s D^2 + V_3 (\io s^3 - s^2) + V_4 s^2 D \bigr).
\end{align*}

As shown in \cite{fouque2011fast}, $f_0$ and $f_1$ are given as follows:
\begin{align*}
f_0(\tau,s) &= \int^{\tau}_0 f_1(u,s)du,\\
f_1(\tau,s) &= \int^{\tau}_0 b(u,s) \ex^{A(\tau,s,u)}du,\\
A(\tau,s,u) &= (\kappa - \rho \sigma \io s + d(s))\frac{1 - g(s)}{d(s)g(s)}\log\Bigl(\frac{g(s) e^{\tau d(s)}-1}{g(s) e^{u \tau d(s)} - 1} \Bigr) + d(s) (\tau - u),\\
d(s) &= \sqrt{(\rho \sigma \io s - \kappa)^2  + \sigma^2(\io s + s^2)},\\
g(s) &= \frac{\kappa - \rho \sigma \io s + d}{\kappa - \rho \sigma \io s - d},\\
b(\tau,s) &= -\bigl( V_1D (-s^2 - \io s ) + V_2 \io s D^2 + V_3 (-\io s^3 + s^2) - V_4 s^2 D \bigr).
\end{align*}
\section{Future work}
As shown here, a correction to the characteristic function of $X_t$, logarithm of the underlying asset price, in the Heston model can be easily obtained in the new model of Fouque and Lorig. Friz et al. \cite{friz2011refined} have proposed asymptotics of the implied volatility in the Heston model which use saddle point approximation for the characteristic function of $X_t$. The next step is to use the corrected characteristic function formula to derive improved implied volatility asymptotics in the corrected Heston model which capture the observed market skew more efficiently.

\section*{Acknowledgements}
I would sincerely like to thank Prof. Jean-Pierre Fouque and Prof. Sandeep Juneja for remarks and helpful discussion.

\bibliographystyle{abbrvnat}
\bibliography{asymptoticHeston}  
\end{document}